\documentclass[a4paper,11pt]{article}
\usepackage{amsmath}
\usepackage{jheppub}
\usepackage{amssymb}
\usepackage{graphicx}
\usepackage[T1]{fontenc}
\usepackage{slashed}
\usepackage[utf8]{inputenc}
\usepackage{xspace}
\usepackage{ulem}
\usepackage{array}
\usepackage{ulem,fancyvrb}
\usepackage{xcolor}

\begin{document}

\title{Cosmic Ray Boosted Sub-GeV Gravitationally Interacting Dark Matter in Direct Detection}

\date{\today}
\author[1]{Wenyu Wang,}
\author[2]{Lei Wu,}
\author[3,4,5]{Jin Min Yang,}
\author[2]{Hang Zhou,}
\author[6,7]{Bin Zhu}

\affiliation[1]{Faculty of Science, Beijing University of Technology, Beijing, China}
\affiliation[2]{Department of Physics and Institute of Theoretical Physics, Nanjing Normal University, Nanjing 210023, China}
\affiliation[3]{CAS Key Laboratory of Theoretical Physics, Institute of Theoretical Physics,
Chinese Academy of Sciences, Beijing 100190, China}
\affiliation[4]{School of Physical Sciences, University of Chinese Academy of Sciences,
Beijing 100049, China}
\affiliation[5]{Department of Physics, Tohoku University, Sendai 980-8578, Japan}
\affiliation[6]{School of Physics, Yantai University, Yantai 264005, China}
\affiliation[7]{Department of Physics, Chung-Ang University, Seoul 06974, Korea}
\emailAdd{wywang@bjut.edu.cn; leiwu@njnu.edu.cn; jmyang@itp.ac.cn; zhouhang@njnu.edu.cn;zhubin@mail.nankai.edu.cn}

\abstract{
Detections of non-gravitational interactions of massive dark matter (DM) with visible sector so far have given null results. The DM may communicate with the ordinary matter only through gravitational interaction. Besides, the majority of traditional direct detections have poor sensitivities for light DM because of the small recoil energy. Thanks to the high energy cosmic rays (CRs), the light DM can be boosted by scattering with CRs and thus may be detected in the ongoing experiments. In this work, we derive the exclusion limits on the cosmic ray boosted sub-GeV DM with gravitational mediator from the Xenon1T data. It turns out that a sizable region of such a cosmic ray boosted DM can be excluded by the current data.
}

\maketitle


\section{Introduction}
The existence of dark matter (DM) in the Universe has been confirmed by astrophysical and cosmological observations. However, the nature of DM is one of the most pressing puzzles of modern physics. Weakly Interacting Massive Particle (WIMP)~\cite{Lee:1977ua} as a compelling dark matter candidate has been searched for in various (in)direct detections~\cite{Jungman:1995df} and collider experiments~\cite{Buchmueller:2017qhf}, most of which aim for DM at the GeV mass scale and above. Recently, the non-observation of WIMPs in those experiments has led to significant efforts focusing on the sub-GeV DM~\cite{Bertone:2018xtm}. Such a light DM is also theoretically motivated and appears in many new physics models~(for recent reviews, see e.g.~\cite{Knapen:2017xzo}), for example, the gravitino~\cite{Pagels:1981ke} and sterile neutrino DM~\cite{Dodelson:1993je}.

As known, the traditional direct detection based on liquid xenon rapidly loses sensitivity for sub-GeV DM, due to the small recoil energy imparted by DM to a nucleus in elastic scattering~\cite{Aprile:2018dbl,Akerib:2017kat,Ren:2018gyx}. The lighter nuclei and lower energy thresholds used in the detectors are able to probe DM in low mass ranges~\cite{Liu:2019kzq,Agnese:2017njq,Angloher:2017sxg,Aguilar-Arevalo:2016ndq}. However, these experiments will lose good discrimination between signal and background events as the DM becomes extremely light. Instead, the searching for DM scattering off electrons may access the lighter DM~\cite{Agnese:2018col,Crisler:2018gci}. Besides, other new methods~\cite{Ibe:2017yqa,Dolan:2017xbu,Bell:2019egg} and new types of detectors~\cite{Hochberg:2015pha,Schutz:2016tid} have been proposed in the past few years.

On the other hand, an observable energy may be imparted to terrestrial nuclear targets by a boosted light DM. There are several acceleration mechanisms of light DM discussed in the literature. Among them, the cosmic ray boosted dark matter (CRDM) is an interesting scenario~\cite{Cappiello:2018hsu,Bringmann:2018cvk}, in which some fraction of the DM halo scattering with the high energy cosmic rays are accelerated to (semi-)relativistic speeds that can produce the detectable scintillation signal in conventional terrestrial experiments~\cite{Cherry:2015oca,Ema:2018bih,Alvey:2019zaa,Cappiello:2019qsw}. For non-relativistic DM, the cross section of DM scattering with nucleons is often assumed to be momentum independent. However, when the mediator mass is lower than the transferred momentum, the full propagator should be included in the scattering cross section to obtain the more accurate results. In Refs.~\cite{Bondarenko:2019vrb,Dent:2019krz}, the sub-GeV CRDM with scalar and vector mediators have been considered in simplified models. Besides, the energetic light CRDM may affect the energy density around $T \sim$ few MeV, and thus is constrained by the BBN data~\cite{Krnjaic:2019dzc}.

In this work, we will focus on a gravitational mediator that couples the light Dirac DM with the SM particles through the energy-momentum tensor. By considering the cosmic ray acceleration mechanism, we will derive the bounds on such a light CDRM with the available direct detection data. This paper is organized as follows: in Sec.~2 we formulate the framework of cosmic ray boosted dark matter in a  simplified DM model with gravitational mediator, in Sec.~3 we present numerical results and discussions, and the conclusion is drawn in Sec.~4.

\section{Model and CRDM}~\label{sec2}
\begin{figure}
  \centering
  \includegraphics[width=8cm]{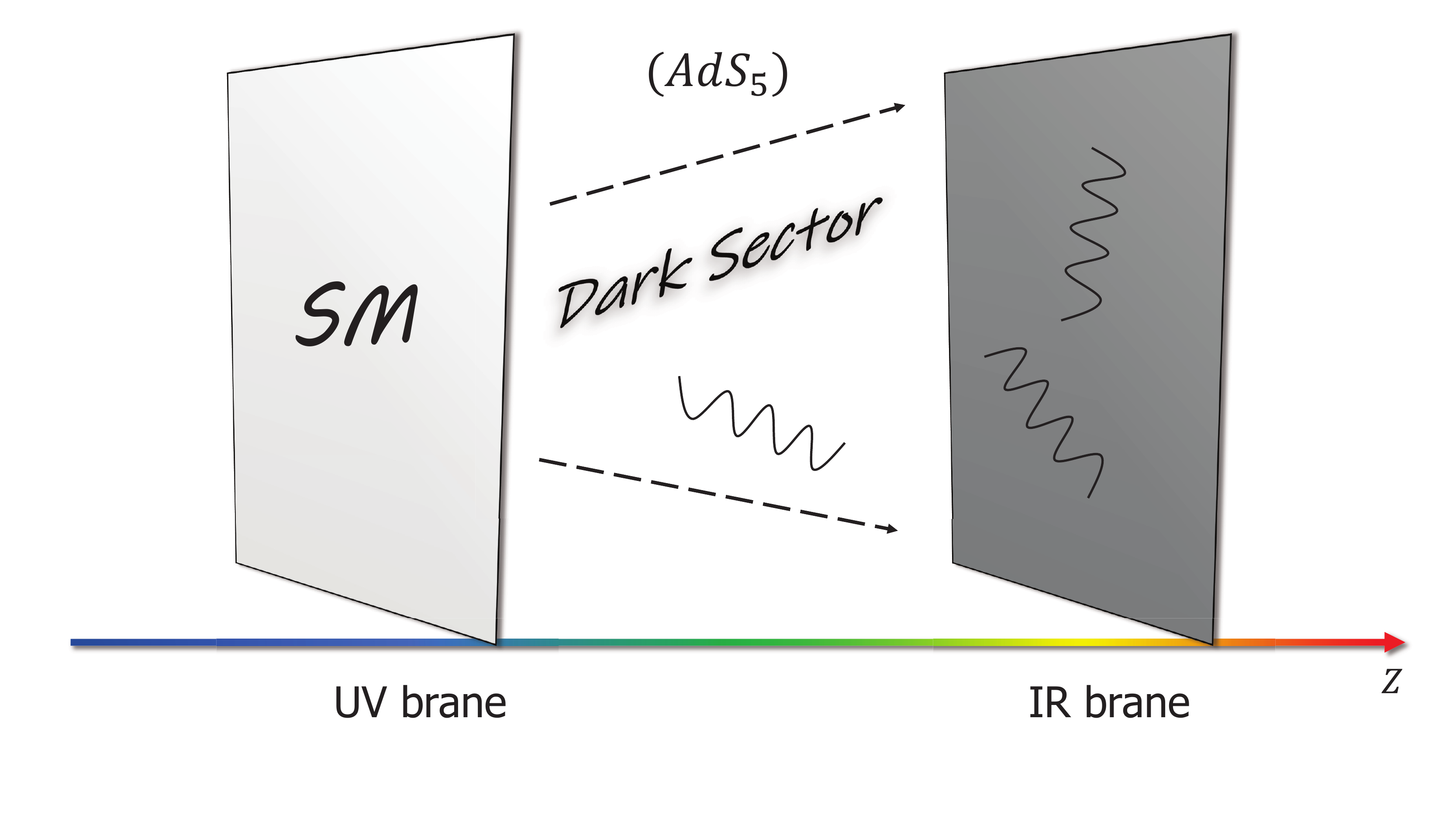}
  \caption{Illustration of gravity-mediated dark matter in warp extra dimension. The SM Higgs boson, electroweak gauge bosons and leptons may live on the UV-brane, while the dark matter, the quarks and gluon may live on the IR-brane ~\cite{Lee:2013bua,Lee:2014caa,Carrillo-Monteverde:2018phy,Brax:2019koq}.}\label{Warped}
\end{figure}

Till now all attempts to detect non-gravitational interactions of DM with ordinary matter have given null results. Thus, it is natural to consider the gravitationally interacting DM (GIDM).  If the mediator is a massless graviton, the UV cut-off is naturally of the Planck scale. Thus the couplings of GIDM with the SM particles are too tiny to produce sizable effects in direct detection. On the other hand, the UV cut-off can be reduced to GeV scale when the spin-2 mediator is a Kaluza-Klein massive graviton, as illustrated in Fig.\ref{Warped}. Such a GIDM scenario can be realized in a warped dark sector~\cite{Lee:2013bua,Lee:2014caa,Carrillo-Monteverde:2018phy,Brax:2019koq}, which is a slice of AdS space in the Ponicare patch with the metric $d s^{2}=(k z)^{-2}\left(\eta_{\mu \nu} x^{\mu} x^{\nu}-d z^{2}\right)$. Here the fifth dimension $z$ is compact and evaluated in the interval $z \in\left[z_{0}, z_{1}\right]$. We mention that $z_0$ is the location of the UV-brane where the SM Higgs boson, electroweak gauge bosons and leptons live, and $z_1$ is the IR-brane where the dark matter, the quarks and gluon live. The IR scale $\Lambda=1/z_1$ is not the electroweak scale since the Higgs boson is now located on the UV brane. Therefore we lose the ability to solve the hierarchy problem, but obtain a choice to probe the dark sector in GeV scale. It thus provides a dynamical mechanism of generating a GeV cut-off scale of GIDM, which can thermally produce the correct relic abundance by gravitational mediators arising from the compactification of extra-dimensions.

Without losing generality, we will perform a model-independent phenomenological study and parameterize the interactions of GIDM and SM fermion as
\begin{eqnarray}
\mathcal{L}_{G}=&-\frac{1}{\Lambda}\left[c_{\mathrm{SM}} G^{\mu\nu}T_{\mu \nu}^{\mathrm{SM}}+c_{\mathrm{DM}} G^{\mu\nu} T_{\mu \nu}^{\mathrm{DM}}\right].
\label{lag}
\end{eqnarray}
where $G^{\mu\nu}$ is the massive $KK$ graviton,  $T_{\mu\nu}^{\mathrm{DM,SM}}$ are the energy-momentum tensors for dark matter and the SM particles, and $\Lambda$ is the inverse of the extra dimension length $\ell$. Such effective interactions allow us to retain the feature of the warped dark sector~\cite{Brax:2019koq}.

With Eq.~\ref{lag}, the tree-level scattering amplitude between fermionic DM and the SM fermion through the spin-$2$ mediator can be written as
\begin{eqnarray}
\mathcal{M} &=& \frac{i c_{\mathrm{DM}}c_{\mathrm{SM}}}{2 m_{G}^{2} \Lambda^{2}}\left(2 \tilde{T}_{\mu \nu}^{\mathrm{DM}} \tilde{T}^{\mathrm{SM}, \mu \nu}-\frac{1}{6} T^{\mathrm{DM}} T^{\mathrm{SM}}\right),
\end{eqnarray}
where $\tilde{T}_{\mu \nu}$ and $T$ are the traceless and trace parts of the energy-momentum tensor. In the momentum space, they are given by
\begin{eqnarray}
\tilde{T}_{\mu \nu}^{q} &=& -\frac{1}{4} \bar{u}_{q}(p_{2})[\gamma_{\mu}(p_{1 \nu}+p_{2 \nu})+\gamma_{\nu}(p_{1 \mu}+p_{2 \mu}) -\frac{1}{2} \eta_{\mu \nu}(\slashed{p}_{1}+\slashed{p}_{2})] u_{q}(p_{1}),\\
T^{q} &=& -\frac{1}{4} \bar{u}_{q}(p_{2})[-6(\slashed{p}_{1}+\slashed{p}_{2})+16 m_{q}] u_{q}(p_{1}),
\end{eqnarray}
where $q$ stands for the fermionic DM or the SM fermions. We present the explicit form of the differential scattering cross section of the DM with the CRs in the appendix.

As mentioned above, the light sub-GeV DM boosted through scattering with the CRs may be accessible in the traditional direct detections. Such a mechanism includes the following steps~\cite{Bringmann:2018cvk}:

{\it Boost of DM by the energetic CRs.} The high energy CRs transfer kinetic energy to the non-relativistic DM particles, making them become energetic flux, which is given by
\begin{eqnarray}
\frac{d \Phi_{\chi}}{d T_{\chi}} &=& D_{\mathrm{eff}} \frac{\rho_{\chi}}{m_{\chi}} \sum_{i} \int_{T_{i}^{\min }} d T_{i} \frac{d \sigma_{\chi i}}{d T_{\chi}} \frac{d \Phi_{i}^{\mathrm{LIS}}}{d T_{i}}.
\label{eqn:flux}
\end{eqnarray}
Here $i$ stands for the specific species of the cosmic rays. We only consider the contributions of $p$ and $^4$He in our calculations. $T_i$ and $T_\chi$ denote the kinetic energy of CRs and DM, respectively. ${d\Phi_i^{\mathrm{LIS}}}/{dT_i}$ is the spectrum of the incoming CR flux taken in the local interstellar (LIS)~\cite{DellaTorre:2016jjf,Boschini:2017fxq}. $\rho_{\chi}$ is the local DM density and $d\sigma_{\chi i}/dT_{\chi}$ is the differential scattering cross section of CR and DM. For simplicity, the source density of CRDM is assumed roughly the same as it is locally within the effective length $D_{\mathrm{eff}} \sim 8$ kpc.

{\it Attenuation of CRDM by the dense matter of the Earth.} When the boosted DM particles travel from the top atmosphere to the location of
detector, the dense matter will degrade the energy of DM~\cite{Starkman:1990nj,Mack:2007xj,Hooper:2018bfw,Emken:2018run}, which can be numerically determined by
\begin{equation}
\frac{d T^z_{\chi}}{d z}=-\sum_{N} n_{N} \int_0^{T_N^{\rm max}}\frac{d \sigma_{\chi N}}{d T_{N}} T_{N} d T_{N}.
\end{equation}
Here $T^z_{\chi}$ is the DM energy at the depth $z$ from the top of the atmosphere. $T_N$ refers to the recoil energy of nucleus $N$. The average density $n_N$ of the Earth's 11 most abundant elements between surface and depth $z$ is calculated by \textsf{DarkSUSY 6}~\cite{Bringmann:2018lay}. ${d\sigma_{\chi N}}/d T_{N}$ is the differential cross section of the CRDM scattering with the dense matter at the depth $z$. Then, the attenuated CRDM flux $d\Phi_{\chi}/dT_{\chi}^z$ at the depth $z$ can be related with the flux ${d\Phi_{\chi}}/{dT_{\chi}}$ at the top of the atmosphere by
\begin{equation}
\frac{d\Phi_{\chi}}{dT_{\chi}^z} =  \left(\frac{dT_\chi}{dT^z_\chi}\right) \frac{d\Phi_{\chi}}{dT_{\chi}}.
\end{equation}

{\it Scattering between the CRDM and ordinary matter in the detector.} In order to derive the bounds on CRDM with the reported limits for heavy DM from conventional direct detections, we define the recoil rate per target particle $N$ mass within the experimentally accessible window of recoil energy $T_1<T_N<T_2$ as
\begin{equation}
R =\int_{T_1}^{T_2} \frac{1}{m_N} d T_{N}\int_{T_{\chi}^{z,\min }}^{\infty}
d T_{\chi}^z \frac{d\sigma_{\chi N}}{dT_N}\frac{d \Phi_{\chi}}{d T_{\chi}^z}.
\label{eq:CRDMrate}
\end{equation}
{The nuclear form factor in case of a gravitational mediator is
derived by using the matching conditions between quarks and nucleons
\cite{Abidin:2009hr}
\begin{equation}
F^2_N(Q^2)=1/(1+Q^2/\Lambda^2_n),
\label{eq:ff}
\end{equation}
which is different from the conventional form ~\cite{Perdrisat:2006hj}.
For the scalar current, i.e. the trace of energy-momentum, we have }
\begin{eqnarray}
\left\langle N(p)\left|T^{\mathrm{SM}}\right| N(p)\right\rangle&=&-m_{N}\left[\sum_{q=u, d, s} f_{T q}^{N}+f_{T G}\right] \bar{u}_{N}(p) u_{N}(p) \nonumber \\
&=& F_s(q^2)m_{N} \bar{u}_{N}(p) u_{N}(p)
\end{eqnarray}
{where $F_s(q^2)$ is the conventional scalar form factor. Its leading order result is $F_s(0)=1$. For the tensor part, i.e. the traceless part of energy-momentum, the gravitational form factor is difficult to obtain. In terms of the AdS/QCD correspondence, we can identify it as
}
\begin{equation}
\tilde{T}_{\mu \nu}^{\chi}\left\langle N\left(p_{2}\right)\left|\tilde{T}^{\psi, \mu \nu}\right| N\left(p_{1}\right)\right\rangle=F_{t}\left(q^{2}\right) \tilde{T}_{\mu \nu}^{\chi} \tilde{T}^{N, \mu \nu}
\end{equation}
with $F_t(0)=-2$.

\section{Numerical results and discussions}~\label{sec3}
\begin{figure}[ht]
\centering
\includegraphics[width=8.5cm]{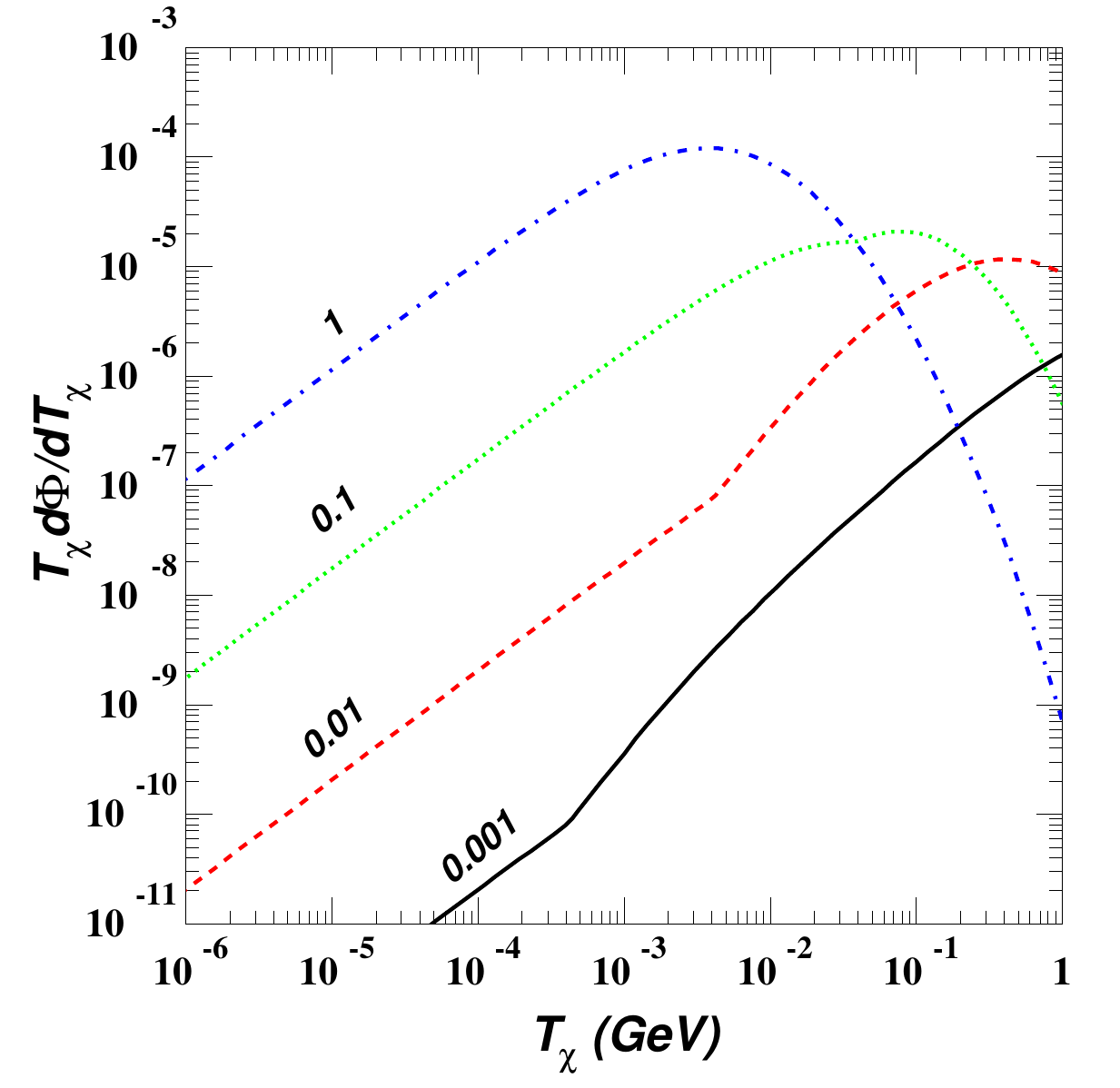}
\caption{The expected flux of CRDM with the spin-2 mediator. The curves from left to right corresponds to DM mass $m_\chi=1, 0.1, 0.01, 0.001$ GeV, respectively. The cut-off scale $\Lambda$ is set at 1 GeV and the couplings $c_{DM}$ and $c_{SM}$ are assumed to be unity.}
\label{fig2}
\end{figure}

In order to include the momentum transfer effect, we use the differential cross section rather than the constant value in our calcuations and thus
modifiy the package \textsf{DarkSUSY}~\cite{Bringmann:2018lay} to numerically calculate the flux of CRDM. The resulting constraint on the DM scattering cross section is obtained by using the Xenon1T experiment. We parameterize the CR flux of protons and helium as in Ref.~\cite{DellaTorre:2016jjf,Boschini:2017fxq}. In Fig.~\ref{fig2}, we show the flux of CRDM with the spin-2 mediator for different DM masses $m_\chi=0.001,0.01,0.1,1$ GeV. From Fig.~\ref{fig2}, we can see that the flux of CRDM has a peak in the (semi-)relativistic velocity region. As expected, the lighter DM particles obtain more kinetic energy through scattering with the CRs. Different from the scalar and vector cases, in the low velocity region the flux increases with the DM mass because the flux is proportional to $m^{2}_{\chi}$ in the limit of $T_\chi \to 0$, as can be seen from Eq.~\ref{eqn:flux} and Eq.~\ref{eq:T0}.

\begin{figure}[ht]
  \centering
  \includegraphics[height=8.5cm,width=8.5cm]{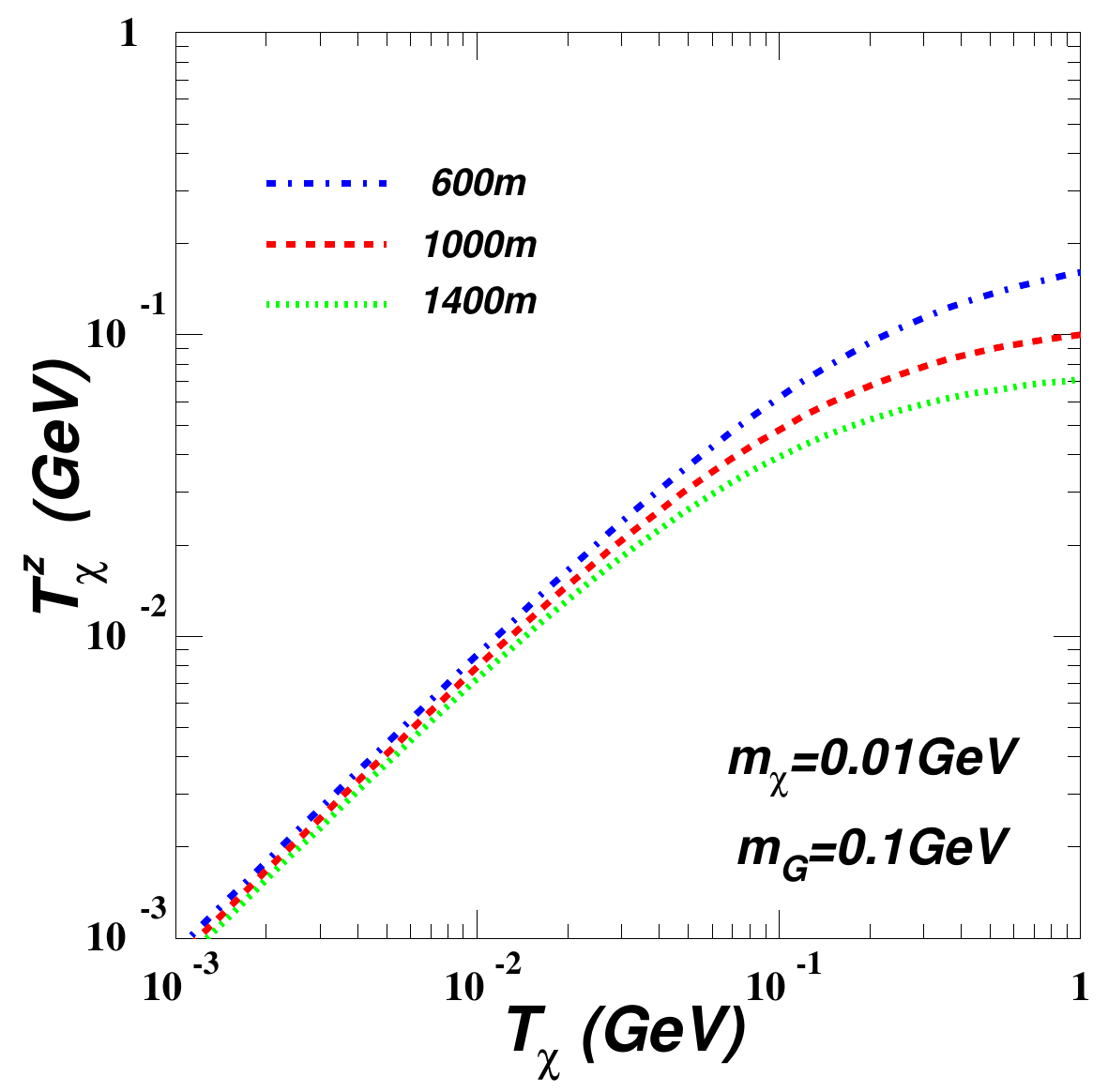}
  \caption{The kinetic energy $T^z_{\chi}$ at different depth of the earth ($z=600,1000,1400~m$) versus the initial energy of CRDM $T_{\chi}$.}\label{attenuation}
\end{figure}
In Fig.~\ref{attenuation}, we display the kinetic energy $T^z_{\chi}$ at different depth of the earth for a benchmark point $m_\chi=0.01$ GeV and $m_\phi= 0.1$ GeV to demonstrate the effect of attenuation. We can see that the scattering between the CRDM and the shield of the earth decelerate the dark matter, and thus $T^z_{\chi}$ decrease with the increase of the depth. Especially when the DM kinetic energy $T_\chi$ is greater than 0.1 GeV, the effects of attenuation can be significant. This indicates that the intensity of the CRDM in the lower energy regions can be augmented.

\begin{figure}[ht]
\centering
\includegraphics[height=8.5cm,width=8.5cm]{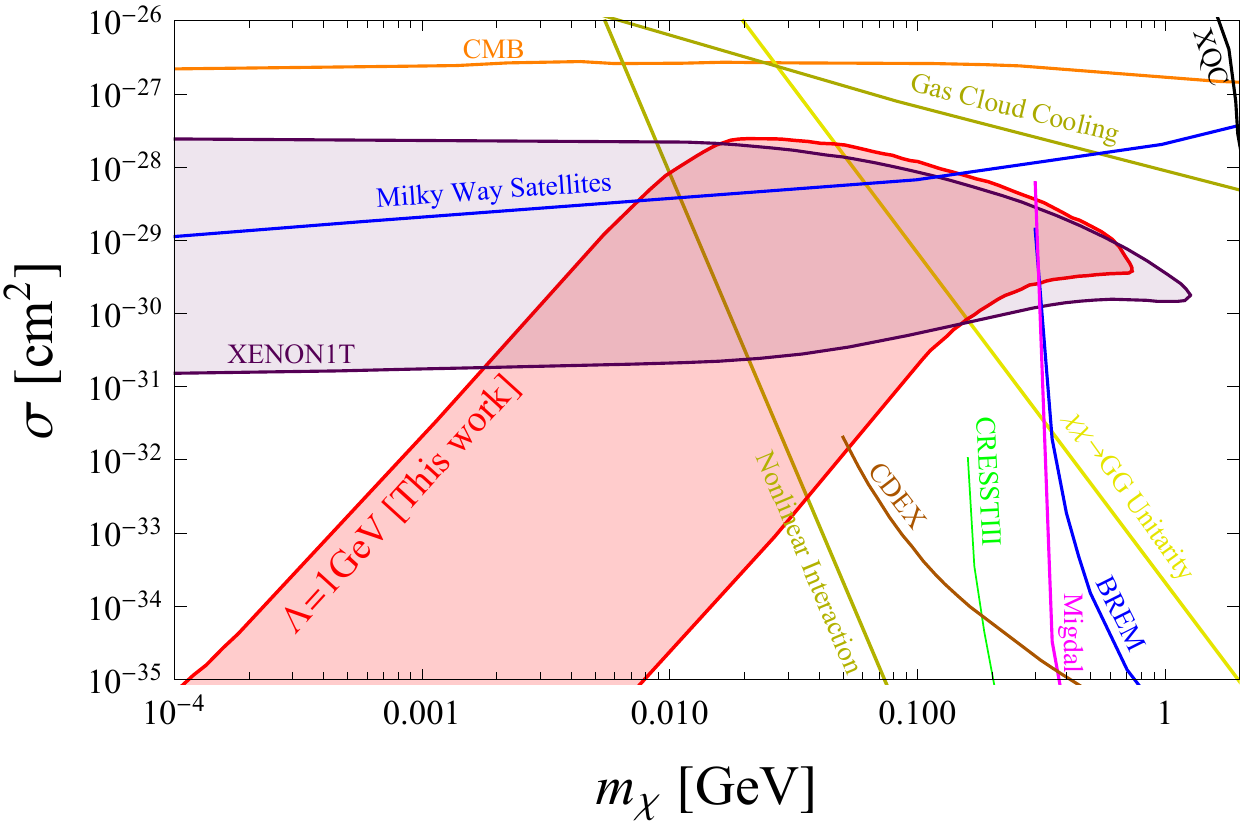}
\caption{Constraints on the spin-independent CRDM cross section from the XENON-1T experiment. We assume the cut-off scale $\Lambda = 1\rm GeV$. The pink color region is the result of this work. The light blue region is the result assuming the constant cross section~\cite{Bringmann:2018cvk}. For comparison, the exclusion curves from the CMB observations~\cite{Xu:2018efh}, the gas cloud cooling~\cite{Bhoonah:2018wmw}, the X-ray Quantum Calorimeter experiment (XQC)~\cite{Mahdawi:2018euy}, other direct detection experiments~\cite{Abdelhameed:2019hmk} and CDEX~\cite{Liu:2019kzq} are also shown.}
\label{fig4}
\end{figure}

{Using the current Xenon 1T data, we obtain the limits on the spin-independent cross section of our CRDM scattering with the nucleons. Our resluts are shown in Fig. \ref{fig4}. For comparison, we also reproduce the result in case of constant cross section given in ~\cite{Bringmann:2018cvk}. We see that the inclusion of the effect of the transferred momentum in the scattering cross section leads to  stronger exclusion limits.  In contrast to the scalar and vector mediators, the heavier DM in low kinetic energy region has a larger flux (as shown in Fig.~\ref{fig2}). This enhances the sensitivity in the heavier DM mass region. Therefore, the exclusion limit on the spin-independent cross section can reach about ${\cal O}(10^{-35})$ cm$^2$ for $10^{-4} {\rm GeV} <m_\chi <10^{-2} {\rm GeV}$. The corresponding mediator mass $m_G$ is constrained to be less than $90\mathrm{MeV}$. Due to the suppression of $\Lambda^{-2}$, we expect that the exclusion limits of the CRDM scattering cross section will weaken with the increase of the cut-off scale $\Lambda$. We need to mention that another path to detect DM with mass below $1\mathrm{GeV}$ is the use of accelerators, such as the MiniBooNE experiment. But due to the current limited sensitivity, the measurement of $\nu-p$ scattering from MiniBooNE gives a weaker bound than Xenon1T.}

{Assuming a constant cross section, we see from Fig. \ref{fig4} that the exclusion region of the cross section seems not sensitive to the dark matter mass. The reason is that the whole detection of CRDM is composed of three steps: 1) the acceleration of the dark matter, 2) the attenuation of the CRDM flux, 3) the detection in the detector. In fact, the interacting cross sections are similar in these three steps. Thus, a large intensity of the flux means a large attenuation of the flux. Then the exclusion region forms a band, as shown in Fig. \ref{fig4}. The survival region below the band is because of the small flux that means not enough dark matter reaching the detector. The survival region above the band has a large intensity of flux, which, however, makes the atenuation stronger so that not enough dark matter can reach the detector. }

{Considering the momentum transfer in the cross section, we see from Fig. \ref{fig4} that the exclusion region of our CRDM can be more sizable in the light dark matter region (below 0.1 GeV). This is because the light dark matter can be easily accelerated to  semi-relativistic or relativistic, and then the nucleus in the detector can have enough recoil energy.  When the mass of dark matter is above 1 GeV,  the acceleration by the cosmic rays can hardly give enough  detection rate, and thus the exclusion ends. The slope of the limits versus the dark matter mass mainly comes from the momentum propagation by the mediator. The effect of momentum propagation can sizably increase the detection rate when the dark matter mass is light. The lighter the dark matter, the more enhancement of the detection rate. Thus smaller cross sections are excluded in case of lighter dark matter. When the dark matter mass approaches 1 GeV, the momentum transfer becomes too weak. Besides, it should be mentioned that the boundary of the excluded region is mainly determined by the mean free path of CRDM in the attenuation process, which is also momentum-dependent.  We numerically calculate the free path for each incident CRDM instead of using a constant value.  }

Finally, we comment on the systematic uncertainties and DM relic density in our scenario.
\begin{itemize}
  \item The main uncertainties arise from the astrophysical inputs, such as the local DM density and CR flux. We assume an NFW profile for the DM distribution~\cite{Navarro:1995iw,Ackermann:2012rg} and a homogeneous CR distribution. We consider the DM within only 1 kpc of the Earth (corresponds to $D_{eff}=0.997$ kpc), which produces limits that are within a factor of 2 of the limits obtained by including the entire CR halo. This will reduce the uncertainties from the shape of the DM density profile.
  \item Several new mechanisms have been proposed to achieve the correct relic density of a thermal sub-GeV DM. Among them, the secluded DM framework~\cite{Pospelov:2007mp}, in which DM interacts with visible sector through a low-mass mediator, can be naturally realized in the warped dark sector by locating the Dirac fermionic DM on the IR brane. The corresponding annihilation cross section is given by,
\begin{equation}
\left<\sigma v_{\mathrm{rel}}\right>_{\chi\chi \rightarrow G G} \simeq \frac{c_{\mathrm{SM}}^{2}c_{\mathrm{DM}}^2 m_{\chi}^{2}}{16 \pi \Lambda^{4}} \frac{(1-r)^{7 / 2}}{r^{2}(2-r)^{2}},
\end{equation}
where $r=m_{G}/m_{\chi}$. Such a process is suppressed by $p$-wave so that it can avoid the constraints from CMB and indirect detections. {When $m_{\chi}$ is smaller than $m_G$, the annihilation cross section is p-wave suppressed, }
\begin{equation}
(\sigma v)_{\chi \bar{\chi} \rightarrow q \bar{q}}=v^{2}  \frac{N_{c} c_{\chi}^{2} c_{q}^{2}}{72 \pi \Lambda^{4}} \frac{m_{\chi}^{6}}{\left(4 m_{\chi}^{2}-m_{G}^{2}\right)^{2}+\Gamma_{G}^{2} m_{G}^{2}}\left(1-\frac{m_{q}^{2}}{m_{\chi}^{2}}\right)^{\frac{3}{2}}\left(3+\frac{2 m_{q}^{2}}{m_{\chi}^{2}}\right)
\end{equation}

 \item {The effective theory is described by two relevant input parameters: $c_q/\Lambda$ and $c_{\chi}/\Lambda$, both of which are constrained by the theoretical consistency and experiments.}  Due to the fact that graviton must be light enough to realize momentum transfer in sub-GeV dark matter, the constraints from meson decay becomes relevant for couplings $c_{\mathrm{SM}}$, $c_{\mathrm{DM}}$ and $\Lambda$. The most severe constraint is the invisible decays of $K^{+}$ or $B^{+}$. The current bounds on the branching ratios are
$\mathrm{BR}\left(K^{+} \rightarrow \pi^{+}+\text {invisible }\right)<\left(1.73_{-1.05}^{+1.15}\right) \times 10^{-10}$\cite{Artamonov:2008qb} and
$\mathrm{BR}\left(B^{+} \rightarrow K^{+}+\text {invisible }\right)<1.6 \times 10^{-5}$. The most dominant decay channel is that a down-type quark $q_1$ decaying into another down-type quark $q_2$ and massive graviton $G$:
\begin{equation}
\Gamma\left(q_{1} \rightarrow q_{2} G\right)=\frac{c_{q}^{2} G_{F}^{2} m_{q_{1}}^{7} u\left(x_{1}\right)}{192(2 \pi)^{5} \Lambda^{2}}\left|\sum_{f=u, c, t} V_{f 1} V_{f 2}^{*} v\left(x_{f}\right)\right|^{2}
\end{equation}
As a result, the bound on quark coupling is derived easily
\begin{equation}
\begin{array}{ll}
\frac{c_{q}}{\Lambda}<0.3 \mathrm{GeV}^{-1}, & K^{+} \rightarrow \pi^{+}+\text {invisible } \\
\frac{c_{q}}{\Lambda}<1.8 \times 10^{-2} \mathrm{GeV}^{-1}, & B^{+} \rightarrow K^{+}+\text {invisible }
\end{array}
\end{equation}
{Thus we can treat $c_{\mathrm{SM}}/\Lambda$ as a phenomenological input parameter whose maximal value is $1.8 \times 10^{-2}\mathrm{GeV}^{-1}$. }

\item {The other relevant parameter of our model is $c_{\chi}/\Lambda$. It can not be constrained by the LHC due to the large QCD background such as the mono-photon or mono-jet. However, we can still constrain it from the theoretical consistency, among which the unitarity puts the strongest constraint:}
\begin{equation}
\left|\mathcal{M}_{\chi \bar{\chi} \rightarrow G G}\right| \simeq \frac{\sqrt{2}}{2} \frac{c_{\chi}^{2} m_{\chi}^{2}}{\Lambda^{2}}\left(\frac{m_{\chi}}{m_{G}}\right)^{2}<8 \pi
\end{equation}
{In addition, the non-linearity of dark matter annihilation puts a constraint on $c_{\chi}/\Lambda$: }
\begin{equation}
\frac{\Lambda}{c_{\chi}} \gtrsim\left(\frac{m_{\chi}^{3}}{m_{G}}\right)^{\frac{1}{2}}
\end{equation}

\item Furthermore, the mono-photon plus leptons at BaBar constrains our model parameters:
\begin{equation}
\begin{aligned}
&\frac{c_{e}}{\Lambda}<2 \times 10^{-4} \mathrm{GeV}^{-1}, \quad \text { BaBar invisible }\\
&\frac{c_{e}}{\Lambda}<3 \times 10^{-5} \mathrm{GeV}^{-1}, \quad \text { BaBar visible }
\end{aligned}
\end{equation}

{Note that the interaction $c_e/\Lambda$ communicates electron with dark matter, which can contribute to the electron recoil. However, our value of $c_e/\Lambda$  is too small to explain the Xenon1T electron recoil excess.}

\end{itemize}

\section{Conclusion}~\label{sec4}
In this paper, we studied the direct detection of the cosmic ray scattering dark matter with a gravitational mediator. Due to the acceleration effect, the sub-GeV CRDM can become (semi-)relativistic and thus be accessible in the conventional direct detections. In contrast with the scalar and vector mediators, the spin-2 mediator produce a larger flux behavior of DM in low energy region due to the nature of tensor interaction, which greatly enhances the sensitivity of heavier DM. By including the momentum-dependent effects, we obtained the exclusion limit of the SI cross section $\sigma_{SI} < {\cal O}(10^{-35})$ cm$^2$ for 0.1 MeV $<m_\chi <$ 10 MeV with the Xenon1T data, which significantly extends the existing limits on such a light DM.

\section*{Acknowledgements}
We especially thanks Hyun Min Lee, Jayden L. Newstead and Peter Athron for helpful discussions. This work is supported by the National Natural Science Foundation of China (NNSFC) under grants No. 11705093, 11775012, 11805161, 11847208, 11875179, 11847612, 11821505, 11675242, 11851303. LW is also supported in part by Jiangsu Specially Appointed Professor Program. JMY is also supported in part by the National Key R\&D Program of China No. 2017YFA0402204. BZ is also supported in part by Natural Science Foundation of Shandong Province under grant No. ZR2018QA007 and by Korea Research Fellowship Program through the National Research Foundation of Korea (NRF) funded by the Ministry of Science, ICT and Future Planning (2019H1D3A1A01070937).

\section*{Appendix}
The differential cross section of the CR scattering with DM in Eq.~\ref{eqn:flux} is given by,
\begin{equation}
\frac{d\sigma_{\chi i}}{dT_{\chi}}=\Delta_0(T_{\chi}^0)+\Delta_1(T_{\chi}^1)+\Delta_2(T_{\chi}^2)+\Delta_3(T_{\chi}^3)+\Delta_4(T_{\chi}^4)
\end{equation}
with
\begin{eqnarray}
\Delta_0(T_{\chi}^0)&=& \frac{m_{\chi } \left(6 T_i m_N+3 T_i^2+2 m_N^2\right)^2}{18 \pi
   \Lambda _c^2 T_i \left(m_G^2+2 m_{\chi } T_{\chi }\right)^2
   \left(T_i+2 m_N\right)} \label{eq:T0} \\
\Delta_1(T_{\chi}^1)&=& -\frac{T_{\chi } \left(6 m_N^3 \left(3 T_i+4 m_{\chi }\right)+8
   m_N^4\right)}{36 \pi  \Lambda _c^2 T_i \left(m_G^2+2 m_{\chi }
   T_{\chi }\right)^2 \left(T_i+2 m_N\right)} \nonumber \\
   &&-\frac{m_N^2 T_{\chi } \left(96 T_i m_{\chi }+9 T_i^2+8 m_{\chi
   }^2\right)}{36 \pi  \Lambda _c^2 T_i \left(m_G^2+2 m_{\chi } T_{\chi
   }\right)^2 \left(T_i+2 m_N\right)} \nonumber \\
   &&- \frac{9 T_{\chi } T_i^2 m_{\chi } \left(4 T_i+m_{\chi }\right)}{36
   \pi  \Lambda _c^2 T_i \left(m_G^2+2 m_{\chi } T_{\chi }\right)^2
   \left(T_i+2 m_N\right)}\\
   &&- \frac{18 T_{\chi } T_i m_N m_{\chi } \left(6 T_i+m_{\chi
   }\right)}{36 \pi  \Lambda _c^2 T_i \left(m_G^2+2 m_{\chi } T_{\chi }\right)^2
   \left(T_i+2 m_N\right)}\\
   \Delta_2(T_{\chi}^2)&=&\frac{2 T_{\chi }^2 m_N^2 (36 T_i+89 m_{\chi })}{288 \pi  \Lambda _c^2 T_i
   \left(m_G^2+2 m_{\chi } T_{\chi }\right)^2 \left(T_i+2 m_N\right)} \nonumber \\
   &&+\frac{36 T_{\chi }^2 m_N^2 m_N
   m_{\chi } (21 T_i+4 m_{\chi })}{288 \pi  \Lambda _c^2 T_i
   \left(m_G^2+2 m_{\chi } T_{\chi }\right)^2 \left(T_i+2 m_N\right)} \nonumber \\
   &&+\frac{ T_{\chi }^2 (9 T_i m_{\chi }(21
   T_i+8 m_{\chi })+72 m_N^3)}{288 \pi  \Lambda _c^2 T_i
   \left(m_G^2+2 m_{\chi } T_{\chi }\right)^2 \left(T_i+2 m_N\right)}\\
   \Delta_3(T_{\chi}^3)&=&-\frac{T_{\chi }^3 m_{\chi } (10 T_i+3 m_{\chi })}{64 \pi  \Lambda _c^2 T_i \left(m_G^2+2
   m_{\chi } T_{\chi }\right)^2 \left(T_i+2 m_N\right)} \nonumber\\
      &&-\frac{T_{\chi }^3 (10 m_N m_{\chi }+3 m_N^2)}{64 \pi  \Lambda _c^2 T_i \left(m_G^2+2
   m_{\chi } T_{\chi }\right)^2 \left(T_i+2 m_N\right)}\\
      \Delta_4(T_{\chi}^4)&=&\frac{m_{\chi } T_{\chi }^4}{64 \pi  \Lambda _c^2 T_i \left(m_G^2+2
   m_{\chi } T_{\chi }\right)^2 \left(T_i+2 m_N\right)}
\end{eqnarray}
where
\begin{equation}
\frac{1}{\Lambda_c^2}=\frac{A^2 F(q^2)c_{\mathrm{SM}}c_{\mathrm{DM}} m_{\chi}^2}{\Lambda^4}
\end{equation}


\end{document}